\documentclass[final,5p,times,twocolumn,12pt]{elsarticle}

\usepackage{amsmath,amsthm,amssymb,url}

\usepackage[nodots]{numcompress}

\journal{Mathematical Biosciences}

\def\b{\beta}

\def\g{\gamma}

\begin{document}

\begin{frontmatter}

\title{Ebola outbreak in West Africa: real-time estimation and multiple-wave prediction}

 \author[label]{Xiang-Sheng Wang\corref{cor}}
 \author[label]{Luoyi Zhong}

\cortext[cor]{\noindent Corresponding author. Email: xswang@semo.edu}

 \address[label]{Department of Mathematics,
 Southeast Missouri State University,
 Cape Girardeau, MO 63701, USA}

\begin{abstract}

Based on the reported data until 18 March 2015 and numerical fitting via a simple formula of cumulative case number, we provide real-time estimation on basic reproduction number, inflection point, peak time and final outbreak size of ongoing Ebola outbreak in West Africa. From our simulation, we conclude that the first wave has passed its inflection point and predict that a second epidemic wave may appear in the near future.

\end{abstract}


\end{frontmatter}

\section{Introduction}

Ebola virus disease (EVD) is a severe disease in humans which has infected nearly 25 thousand individuals and claimed more than ten thousand deaths during the recent outbreak in West Africa, according to the report of World Health Organization dated 18 March 2015 \cite{WHO,CDC}. The most affected countries are Guinea, Liberia and Sierra Leone. This study aims to provide some real-time estimations on the outbreak in these three countries using the reported cumulative case data.
Specifically, we will estimate the following quantities:
\begin{enumerate}
  \item basic reproduction number $R_0$, which is defined as the average new cases caused by a single infective individual during one infectious period;
  \item inflection point $t_c$, which marks the time when the increment speed of cumulative case numbers starts to slow down;
  \item final outbreak size $K$, which indicates the total number of infectious cases throughout the outbreak wave.
  \item peak time $t_p$, which is defined as the critical time when daily infectious number reaches its maximum.
\end{enumerate}
All of these indicators provide quantitative information about severity of a disease outbreak.

\section{Methods}
Following \cite{WWY12}, we study the epidemic model:
\begin{equation}\label{model}
  \begin{aligned}
    S'(t)&=-{\beta S(t)I(t)\over S(t)+I(t)};\\
    I'(t)&={\beta S(t)I(t)\over S(t)+I(t)}-\gamma I(t),
  \end{aligned}
\end{equation}
where $S(t)$ and $I(t)$ are the numbers of susceptible and infective individuals at time $t$, respectively.
The constant $\beta$ denotes the transmission rate of the disease,
and the constant $\gamma$ corresponds to the removal rate of infective individuals.
The basic reproduction number \cite{DHR90,vW02} is given by
\begin{equation}\label{R0}
  R_0={\beta\over\gamma}.
\end{equation}
It is noted that a disease outbreak occurs if and only if $R_0>1$.
The differential system \eqref{model} can be solved explicitly and its solution is given by
\begin{equation}\label{SI}
  \begin{aligned}
    S(t)=&K[1+e^{\gamma(R_0-1)(t-t_c)}]^{-R_0/(R_0-1)};\\
    I(t)=&K[1+e^{\gamma(R_0-1)(t-t_c)}]^{-1/(R_0-1)}
    \\&-K[1+e^{\gamma(R_0-1)(t-t_c)}]^{-R_0/(R_0-1)},
  \end{aligned}
\end{equation}
where $K$ and $t_c$ are two constants of integration.
Now, we define the cumulative infective case number at time $t$ as
\begin{equation}
  C(t)=\int_{-\infty}^t{\beta S(t)I(t)\over S(t)+I(t)}.
\end{equation}
From \eqref{model} and \eqref{SI}, we have
\begin{equation}\label{C}
  C(t)=K-K[1+e^{\gamma(R_0-1)(t-t_c)}]^{-R_0/(R_0-1)}.
\end{equation}
Here, the constant $K=C(\infty)$ has the biological meaning of final outbreak size.
It can be verified that $C''(t_c)=0$. Hence, $t_c$ is the inflection point of $C(t)$.
We remark that the inflection point $t_c$ is related to but different from another commonly used quantity: the peak time, denoted by $t_p$.
The peak time is defined as the time when infective case number achieves its maximum, namely, $I'(t_p)=0$.
It follows from \eqref{SI} that
\begin{equation}
  t_p=t_c+{\ln R_0\over\gamma(R_0-1)}.
\end{equation}
In the case when $R_0$ is close to $1$, namely, $\ln R_0\approx R_0-1$, we can approximate the difference $t_p-t_c$ by $1/\gamma$.
Thus, the peak time occurs about one infectious period after the inflection point \cite{WWY12}.

Richards' empirical model \cite{Ri59} was suggested to provide real-time estimation of a disease outbreak; see \cite{HC06} for example.
However, some of the parameters in Richards' model do not have clear biological meanings \cite{WWY12}.
The advantage of formula \eqref{C} is that all of the parameters in this formula have significant biological interpretations.
We will use the explicit formula of $C(t)$ in \eqref{C} to fit the reported cumulative case numbers of 2014 Ebola outbreak in West Africa and provide real-time estimation of basic reproduction number $R_0$, inflection point $t_c$, final outbreak size $K$ and peak time $t_p$.

As pointed out in \cite{WWY12}, one should fix the value of $\gamma$, the removal rate of infective individuals, to resolve possible overfitting problems.
Note that $1/\gamma$ can be regarded as the infectious period which characterizes the average duration of an individual being infective.
In most cases, an individual is removed from the infective group either by recovery or death.
For the fatal cases of Ebola virus disease, death usually occurs between 6 and 16 days (with mean 7.5 days) after onset of symptom; and for the non-fatal cases, patients may improve their symptoms at around day 6 but need more time to recover \cite{CDC2}.
Convalescent patients may still be infective because the Ebola virus RNA may remain in the body fluid for a couple of weeks even though the risk of transmission from them is low \cite{BTD07}.
It is thus reasonable to assume the infectious period to be 7.5 days with some possible perturbations in the interval between 6 and 16 days.
In our simulation, we first fix $1/\gamma=7.5$ days to estimate the basic reproduction number $R_0$, inflection point $t_c$, peak time $t_p$ and final outbreak size $K$ using reported cumulative case data of Ebola virus in Guinea, Liberia and Sierra Leone, respectively \cite{WHO,CDC}.
Also, we provide the 95\% confidence intervals of each estimated parameter value using bootstrap method.
Next, we vary the value of the parameter $1/\gamma$ from 6 to 16 days and investigate the sensitivity of fitted parameter values.

\section{Results}

The basic reproduction number $R_0$ is estimated as 1.116 (95\% CI: 1.115-1.116) for Guinea, 1.226 (95\% CI: 1.225-1.228) for Liberia, and 1.181 (95\% CI: 1.181-1.182) for Sierra Leone. The inflection point $t_c$ is estimated as 21 November 2014 for Guinea, 24 October 2014 for Liberia, and 28 November 2014 for Sierra Leone. As shown in Table \ref{table-fit}, the lengths of 95\% confidence intervals for the estimated inflection points are no more than one day.

\begin{table*}[htp]
\centering
\begin{tabular}{ccccc}
\hline
& final outbreak size $K$ & basic reproduction number $R_0$ & inflection point $t_c$ & peak time $t_p$\\
\hline
Guinea & 3268 [3257, 3274] & 1.116 [1.115, 1.116] & 266 [265, 266] & 273 [272, 273]\\
Liberia & 8630 [8605, 8660] & 1.226 [1.225, 1.228] & 238 [237, 238] & 244 [244, 245]\\
Sierra Leone & 11227 [11198, 11253] & 1.181 [1.181, 1.182] & 273 [272, 273] & 279 [279, 280]\\
\hline
\end{tabular}
\caption{Estimated parameter values with 95\% confidence intervals. Here, day 1 corresponds to 1 March 2014. So, days 266, 238 and 273 correspond to 21 November 2014, 24 October 2014, and 28 November 2014, respectively.}
\label{table-fit}
\end{table*}

\begin{figure}[h]
\centering
\includegraphics[width=\linewidth]{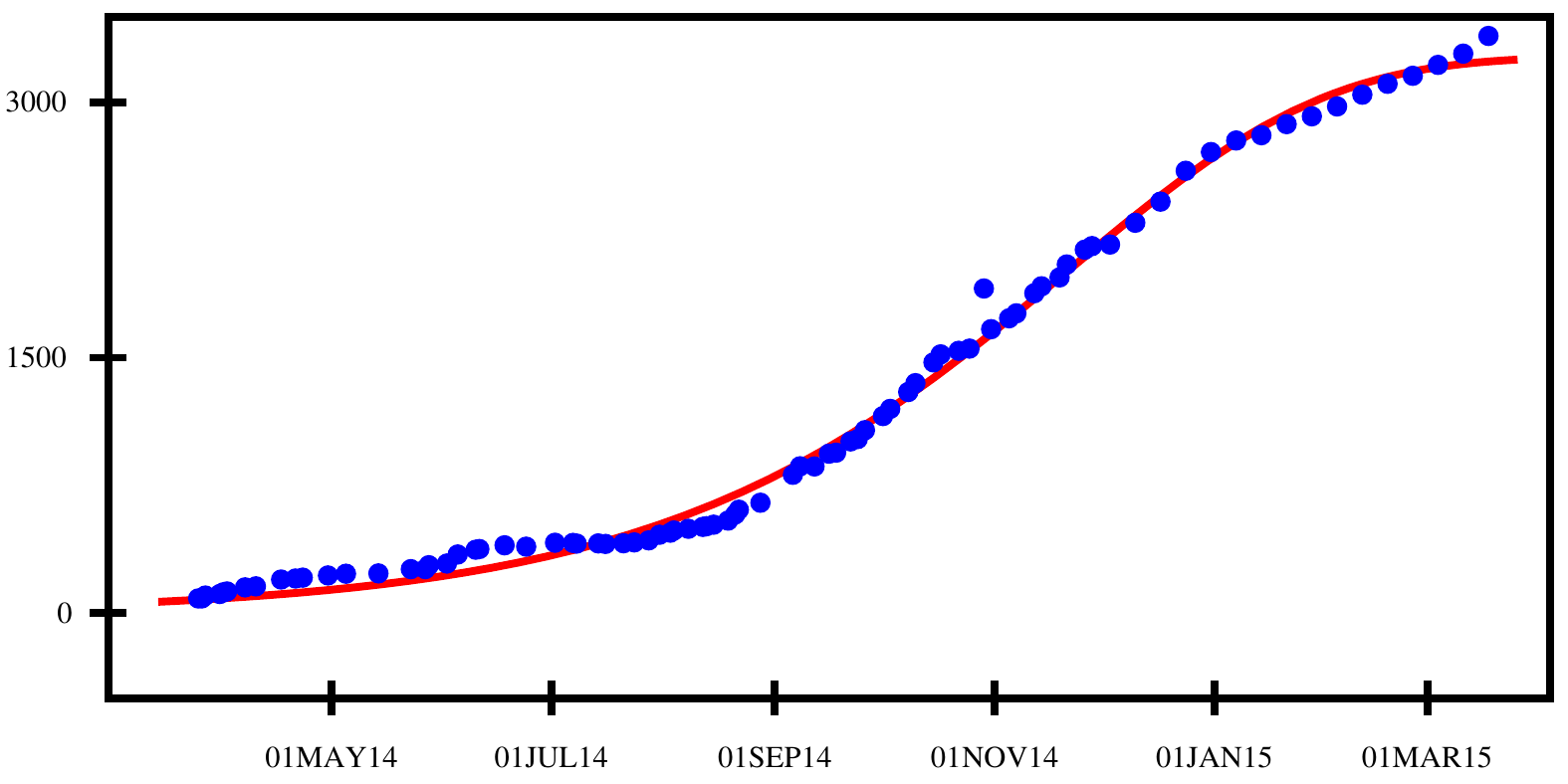}
\caption[Guinea]{Fitted graph for the reported cumulative cases in Guinea. The dots are real data and the curve is plotted using fitted results. }\label{fig-Guinea}
\end{figure}

\begin{figure}[h]
\centering
\includegraphics[width=\linewidth]{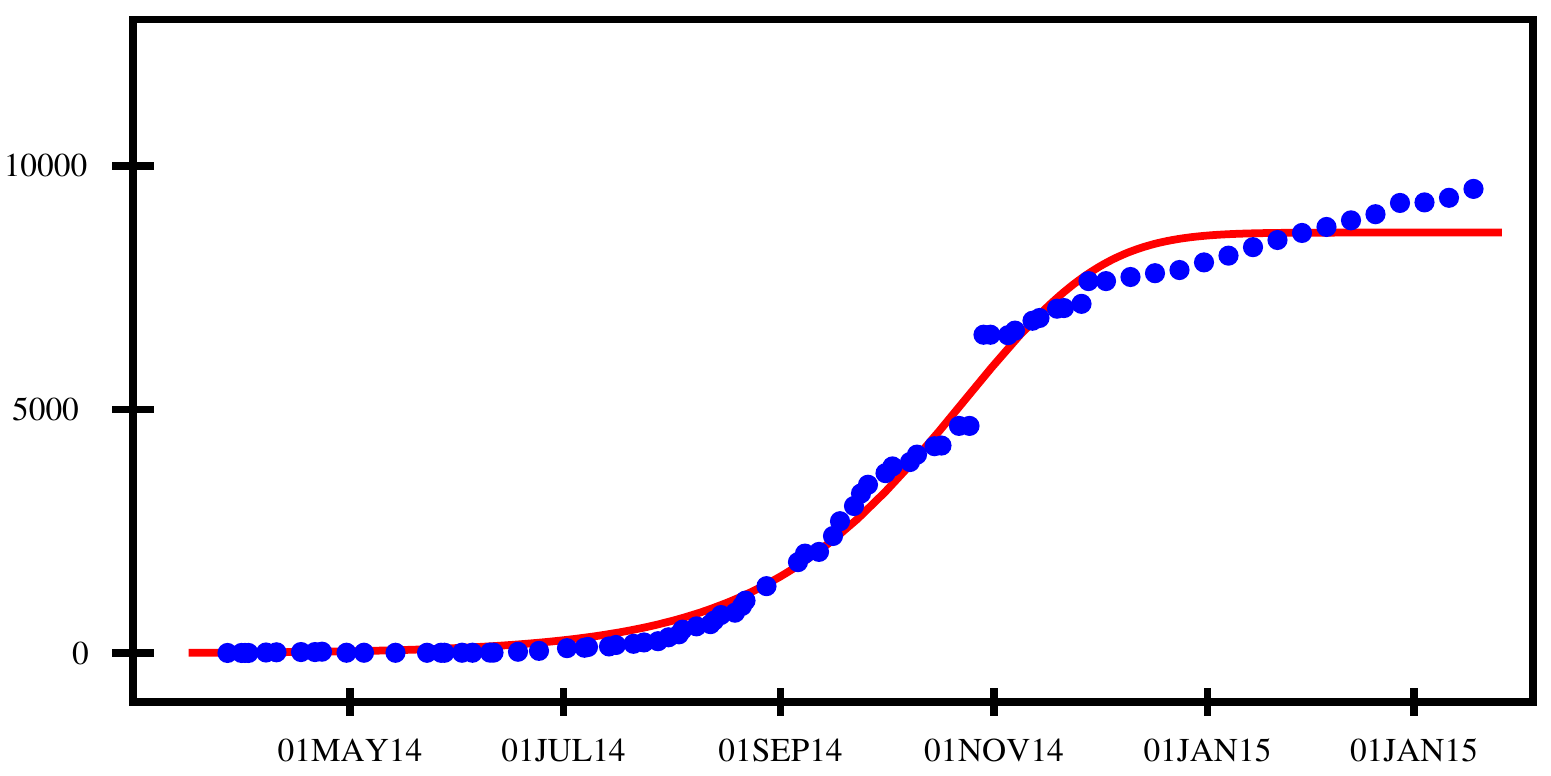}
\caption[Guinea]{Fitted graph for the reported cumulative cases in Liberia. The dots are real data and the curve is plotted using fitted results. }\label{fig-Liberia}
\end{figure}

\begin{figure}[h]
\centering
\includegraphics[width=\linewidth]{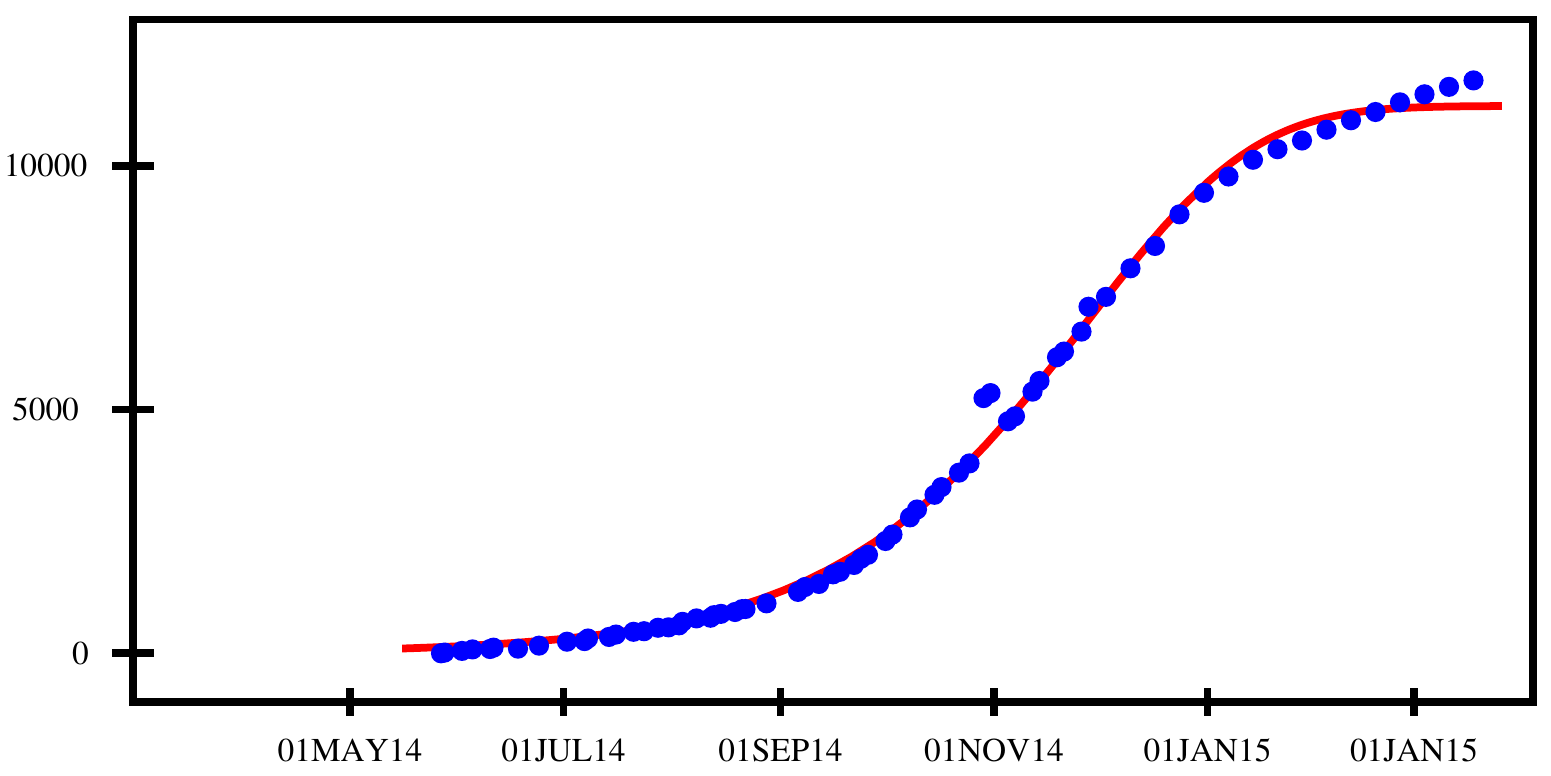}
\caption[Guinea]{Fitted graph for the reported cumulative cases in Sierra Leone. The dots are real data and the curve is plotted using fitted results. }\label{fig-SierraLeone}
\end{figure}

The fitted curves together with reported cumulative case data are illustrated in Figure \ref{fig-Guinea} (Guinea), Figure \ref{fig-Liberia} (Liberia) and Figure \ref{fig-SierraLeone} (Sierra Leone). It is noted that in each of these three figures, there is a jump on the reported cumulative case numbers in late October 2014. This is due to a more comprehensive assessment of patient databases on the World Health Organization report dated 29 October 2014 \cite{WHO2}. Among these three countries, Liberia has the most significant gap, which may account for the result that the inflection point for Liberia is about one month earlier than the other two countries.

We also fit the final outbreak size as 3268 for Guinea, 8630 for Liberia, and 11227 for Sierra Leone. All of these estimated values are smaller than cumulative case numbers reported on 18 March 2015. This indicates that another potential outbreak wave may be approaching \cite{HC06}.

\begin{figure}[h]
\centering
\includegraphics[width=\linewidth]{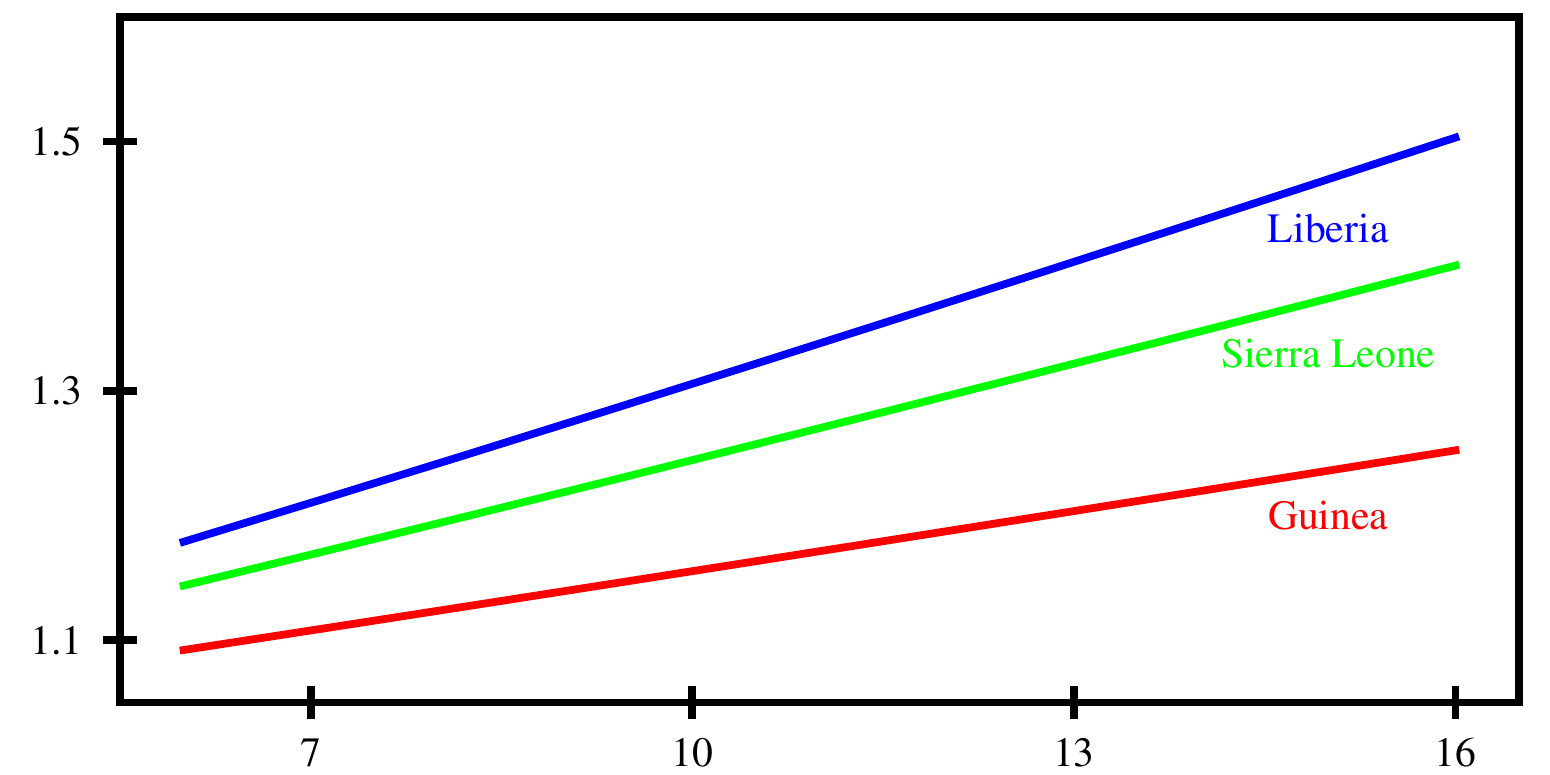}
\caption[Guinea]{Estimated values of basic reproduction number when infectious period increases from 6 to 16 days.}\label{fig-R0}
\end{figure}

\begin{figure}[h]
\centering
\includegraphics[width=\linewidth]{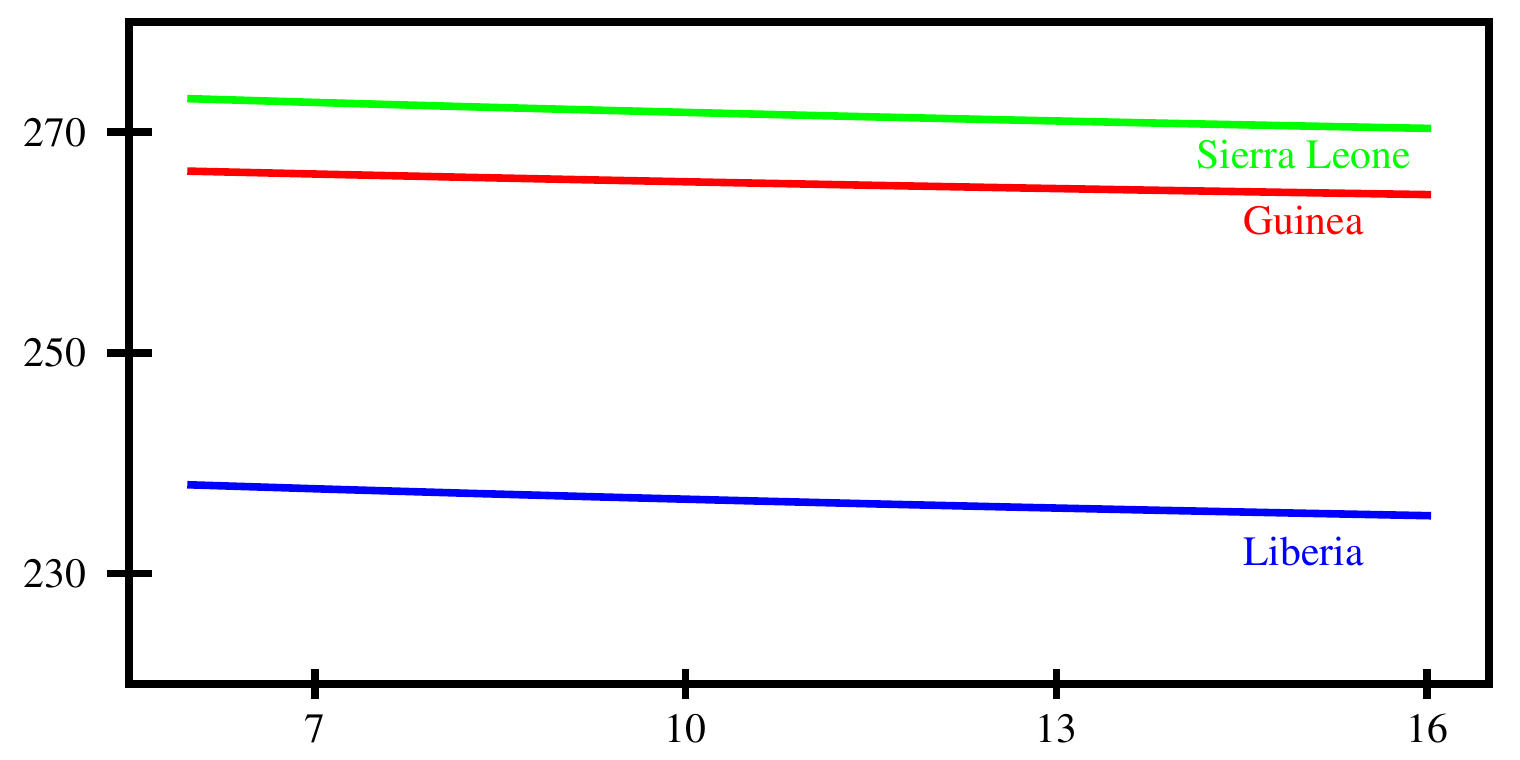}
\caption[Guinea]{Estimated values of inflection point when infectious period increases from 6 to 16 days. Here, day 1 corresponds to 1 March 2014. So, days 266, 238 and 273 correspond to 21 November 2014, 24 October 2014, and 28 November 2014, respectively.}\label{fig-tc}
\end{figure}

\begin{figure}[h]
\centering
\includegraphics[width=\linewidth]{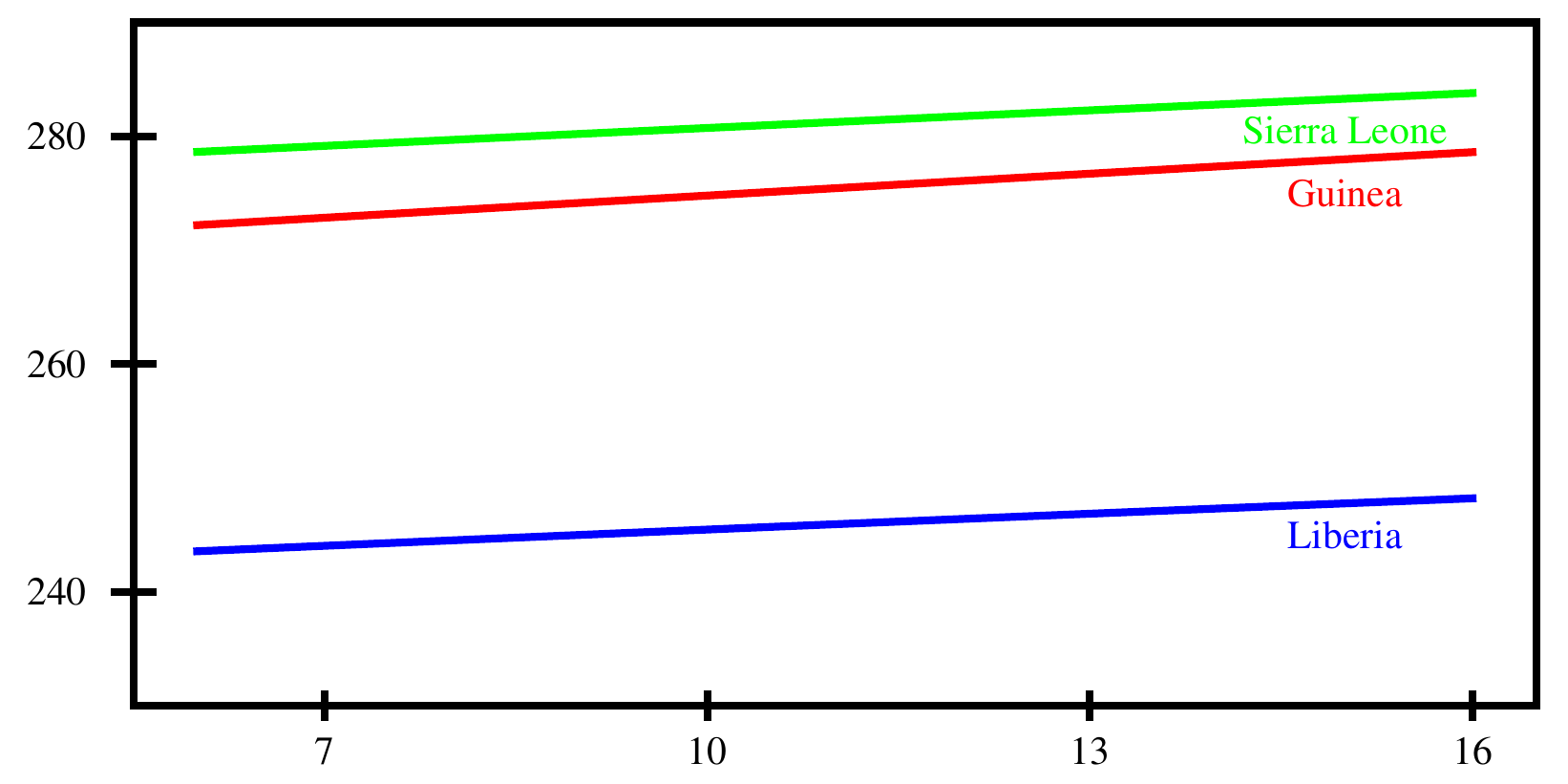}
\caption[Guinea]{Estimated values of peak time when infectious period increases from 6 to 16 days. Here, day 1 corresponds to 1 March 2014. So, days 266, 238 and 273 correspond to 21 November 2014, 24 October 2014, and 28 November 2014, respectively.}\label{fig-tp}
\end{figure}

\begin{figure}[h]
\centering
\includegraphics[width=\linewidth]{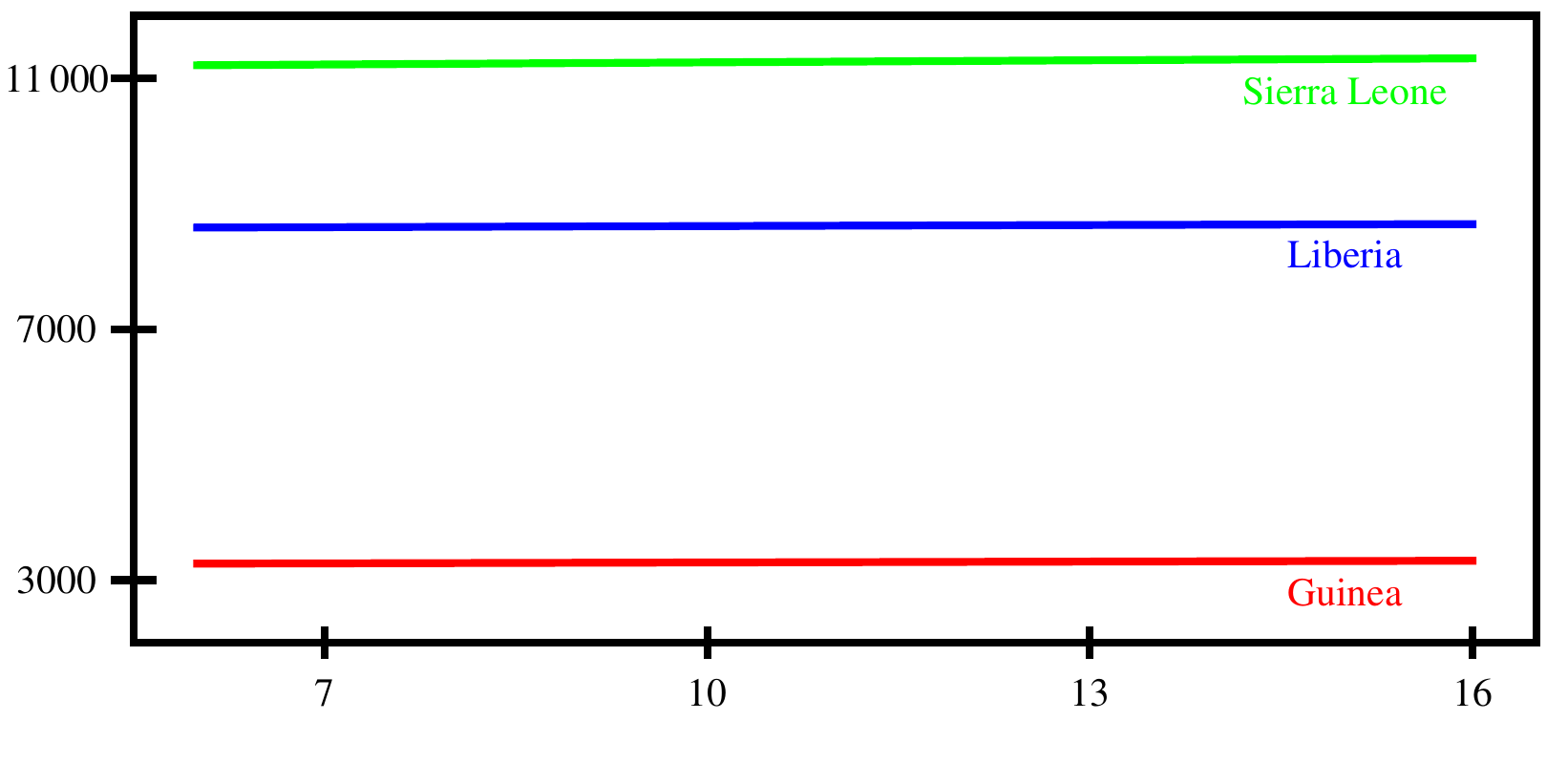}
\caption[Guinea]{Estimated values of final outbreak size when infectious period increases from 6 to 16 days.}\label{fig-K}
\end{figure}

Now, we regularly increase the value of infectious period $1/\gamma$ from 6 to 16 days, and conduct numerical simulations.
It is noted that the fitted values of basic reproduction number $R_0$ will also increase from 1.092 to 1.252 for Guinea, from 1.179 to 1.503 for Liberia, and from 1.144 to 1.400 for Sierra Leone; see Figure \ref{fig-R0}.
The estimated final outbreak size stays in a range of [3261, 3307] for Guinea, [8620, 8677] for Liberia, and [11209, 11319] for Sierra Leone; see Figure \ref{fig-K}.
On the other hand, the inflection point $t_c$ and peak time $t_p$ do not vary too much; see Figures \ref{fig-tc} and \ref{fig-tp}. For Guinea, $t_c$ decreases from 266 (21 November 2014) to 264 (19 November 2014). For Liberia, $t_c$ decreases from 238 (24 October 2014) to 235 (21 October 2014). For Sierra Leone, $t_c$ decreases from 273 (28 November 2014) to 270 (25 November 2014). The peak time $t_p$ increases from 272 (27 November 2014) to 279 (4 December 2014) for Guinea, from 244 (30 October 2014) to 248 (3 November 2014) for Liberia, and from 279 (4 December 2014) to 284 (9 December 2014) for Sierra Leone.
We observe that the fitted inflection point $t_c$ and peak time $t_p$ are stable under perturbations on the infectious period $1/\gamma$.

\section{Discussion}
This study provides real-time estimation of basic reproduction number, final outbreak size, inflection point and peak time for the onging Ebola outbreak in West Africa using reported cumulative case data.

The fitted basic reproduction numbers are smaller than those estimated in \cite{Al14} where only the data until 20 August 2014 was used.
This indicates that the disease control policy became more effective during the late stage of the outbreak.

We also observe that the increment speed of cumulative case number began to slow down after 21 November 2014 in Guinea, 24 October 2014 in Liberia, and 28 November 2014 in Sierra Leone. The estimated inflection points for Guinea and Sierra Leone are close to each other, but the one for Liberia is about one month earlier. This is due to a significant increase on reported cumulative case number dated 29 October 2014; see Figure \ref{fig-Liberia} and \cite{WHO2}.

From Table \ref{table-fit}, we note that the estimated peak time has about one week's delay after the estimated inflection point, while the infectious period is fixed as 7.5 days.
This supports the conclusion in \cite{WWY12} that the peak time occurs about one infectious period after the inflection point.

If we vary the infectious period from 6 to 16 days, the estimated basic reproduction number stays in a range of [1.092, 1.252] for Guinea, [1.179, 1.503] for Liberia, and [1.144, 1.400] for Sierra Leone. The estimated final outbreak size ranges from 3261 to 3307 for Guinea, from 8620 to 8677 for Liberia, and from 11209 to 11319 for Sierra Leone. The estimated inflection point and peak time are much stabler and only varies within a small interval.
This demonstrates that our method has a significant accuracy in capturing the inflection point and peak time.

The values of final outbreak sizes in three countries are all underestimated, which can be considered as a warning signal of a second outbreak wave.

\appendix

\section{Explicit solution of \eqref{model}}
Here, we provide the detail in solving the system \eqref{model}.
First, we add the two equations in \eqref{model} to obtain
$S'(t)+I'(t)=-\g I(t)$.
Coupling this with the first equation of \eqref{model} yields
$$\frac{d(S+I)}{dS}=\frac{\g(S+I)}{\b S}.$$
This equation is separable and its solution can be written as
$$C_1(S+I)=S^{\g/\b},$$
where $C_1>0$ is a constant of integration. 
Next, we use the above relation to eliminate $I$ in the first equtaion of \eqref{model}.
It follows that
$$S'(t)=-\b S(t)+\b C_1[S(t)]^{2-\g/\b}.$$
This is a Bernoulli equation. We set $u(t)=[S(t)]^{\g/\b-1}$ and obtain
$$u'(t)=(\b-\g)[u(t)-C_1].$$
The above equation becomes separable and its solution is given by
$$u(t)=C_1[1+C_2e^{(\b-\g)t}]$$
with $C_2>0$ being another constant of integration.
Now, we have
$$S(t)=C_1^{\b/(\g-\b)}[1+C_2e^{(\b-\g)t}]^{\b/(\g-\b)}.$$
Substituting this into the relation between $S$ and $I$ gives
\begin{align*}
I(t)=&C_1^{\b/(\g-\b)}[1+C_2e^{(\b-\g)t}]^{\g/(\g-\b)}
\\&-C_1^{\b/(\g-\b)}[1+C_2e^{(\b-\g)t}]^{\b/(\g-\b)}.
\end{align*}
Finally, we set $K=C_1^{\b/(\g-\b)}$ and $t_c=-(\ln C_2)/(\b-\g)$ to rewrite the formulas of $S(t)$ and $I(t)$ as
\begin{align*}
S(t)=&K[1+e^{(\b-\g)(t-t_c)}]^{\b/(\g-\b)};\\
I(t)=&K[1+e^{(\b-\g)(t-t_c)}]^{\g/(\g-\b)}
\\&-K[1+e^{(\b-\g)(t-t_c)}]^{\b/(\g-\b)}.
\end{align*}
This is equivalent with \eqref{SI} in view of \eqref{R0}.

\end{document}